\documentclass[fleqn,10pt]{SelfArx} 

\definecolor{color1}{RGB}{0,0,90} 
\definecolor{color2}{RGB}{0,20,20} 

\newlength{\tocsep} 
\usepackage{lipsum} 

\usepackage{url}
\usepackage{gb4e}

\JournalInfo{January 22, 2014 (11/7/2023 rev)} 
\Archive{PARC} 

\PaperTitle{Difference Based Content Networking} 

\Authors{Marc Mosko\textsuperscript{1}*} 
\affiliation{\textsuperscript{1}\textit{Palo Alto Research Center}} 
\affiliation{*\textbf{Corresponding author}: marc.mosko@parc.com} 

\newcommand{\keywordname}{Keywords} 

\Abstract{
Difference Based Content Networking (DBCN) uses diffs between versions to optimize the amount
of data sent on the network for version updates of data files.   This paper describes several variations
on DBCN based on binary diffs, content object diffs, byte offset diffs, and chunk catalog diffs.
Existing CCNx versioning methods use total replication of data file bytes between versions, resulting
in poor efficiency in data transfer.
}

\begin{document}

\flushbottom 

\maketitle 

\tableofcontents 

\thispagestyle{empty} 


\section*{Introduction} 

\addcontentsline{toc}{section}{\hspace*{-\tocsep}Introduction} 

Difference Based Networking is an ICN approach to distributing patched objects.  ICN protocols, such 
as CCNx~\cite{rfc8569,rfc8609} name chunks of data that constitute an application object.  Typically,
a manifest~\cite{irtf-icnrg-flic-05} enumerates the hash-based name for each chunk of data.  The form
of the manifest is not important, and we use a more generic name of Aggregated Signing Objects (ASOs).
The name ASO comes from the fact that a single object carrying a cryptographic signature can aggregate the
signing of all the subsequent chunks because they are named by a cryptographically secure hash name.

Diff-based content networking (DBCN) uses differences between versions to reduce the amount of data transferred on the network. 
For example, if a user updates a 10 Mbyte file with a 1 Kbyte change, it would be encoded as the original 10 Mbyte file and
a second 1 Kbyte diff.  If a remote user already has the 10 Mbyte file, she would only need to transfer the 1 Kbyte diff.

DBCN is similar to versioned file systems, or in some ways journaled file systems.  There is a ground truth of the original object, then
 a series of diffs.  At some point, a new ground truth may be written to avoid needing a large number of diffs.  Different implementations
 of DBCN may use different strategies for writing new ground truths or consolidating diffs to optimize object transfer.
 
DBCN uses secure catalogs, sometimes called Aggregated Signing Objects (ASOs).  This allows a user to limit cryptographic signing to only the secure catalog (ASOs) rather than sign every content object that constitute the underlying data.
 
 DBCN exploits the use of secure catalogs to efficiently encode version differences by referencing the secure catalog
 of an earlier version and then indicating the differences to the new version.   
 There are many different ways of using diff-based encoding.  This article describes several alternatives, including content diffs and ASO diffs.
In the following, we describe using a ``binary diff'' at times.  This means measuring the byte location of a difference and indicating the
new bytes that should replace the old bytes.  For text-based data, a standard text diff could be used instead, with the proper
indication of the diff type in the encoding.

The preferred method of DBCN is described in Section~\ref{sec:diffchunks}.  This method combines data reduplication
techniques with differentially-encoded secure catalogs.  Earlier sections describe other variations the provide background
as to why Section~\ref{sec:diffchunks} is the preferred method. 

Unless otherwise indicated, a secure catalog does not need to reference the immediately preceding secure catalog
to form the secure catalog tree.  It may skip back to earlier versions, or not include one at all.


\section{Diff-encoded content objects}
\begin{figure}
\centering
\includegraphics[width=\linewidth, clip, page=1, trim=0in 0in 0in 2.5in]{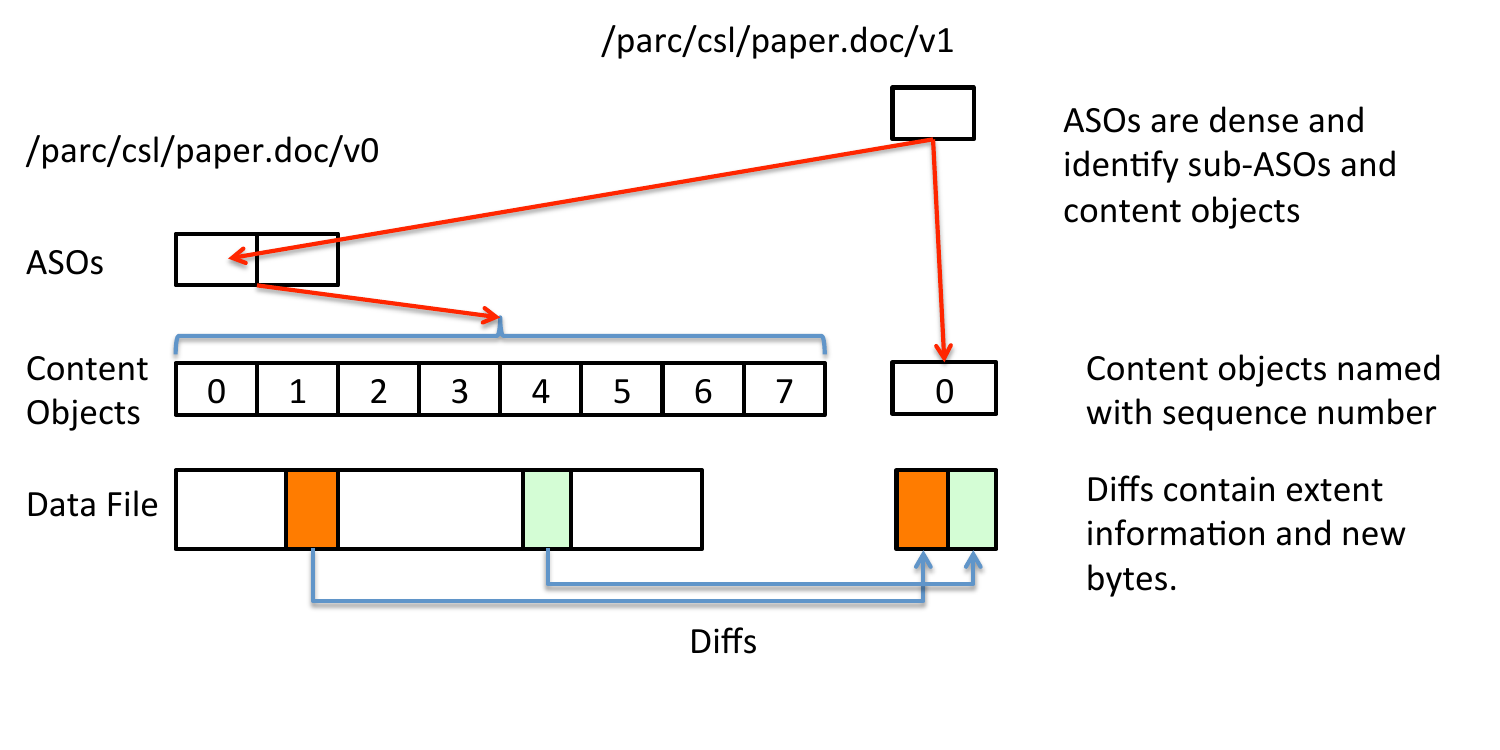}
\caption{Differential data using binary diff encoding}
\label{fig:basicdiff}
\end{figure}

In the first variation of diff-encoded content objects, as shown in Fig.~\ref{fig:basicdiff},
a data file (or memory) is partitioned into a set of content objects and those
content objects are named with sequential numbers.  This is a traditional way to generate
CCNx content objects from an underlying data file.  

Each content object would be named with its sequence number, such as \url{/parc/csl/papers.doc/v0/s0},
where \url{s0} indicates segment 0 of version 0.  Each content object has an implicit Content Object Hash,
being the SHA-256 cryptographic hash of the content object.  This allows exact retrieval of a
matching content object with cryptographic verification that the object received is the object desired.

The secure catalog would have a name such as \url{/parc/csl/paper.doc/v0}, where
the component \url{v0} indicates the first version.   The catalog enumerates the Content Object Hash
of each constituent content object in order.  This allows complete reconstruction of the data file
using only the signature on the secure catalog.

A new version, such as  \url{/parc/csl/paper.doc/v1}, has a secure catalog that points to the previous
version ( \url{/parc/csl/paper.doc/v0}) and the new content objects that make up the binary diff.  The
binary diff is a structured list of the binary differences from \url{v0} to \url{v1}, such that one could
``patch'' the older version to get the newer version.  Due to insertions or deletions in the new version,
this may be an inefficient representation.

In this method, the encoded content objects only contain the minimum number of difference bytes between
the two versions and annotations inside the content object payload describe where those differences
occur in the previous version.

\section{Content Object diffs}
\begin{figure}
\centering
\includegraphics[width=\linewidth, clip, page=2, trim=0in 0in 0in 2.5in]{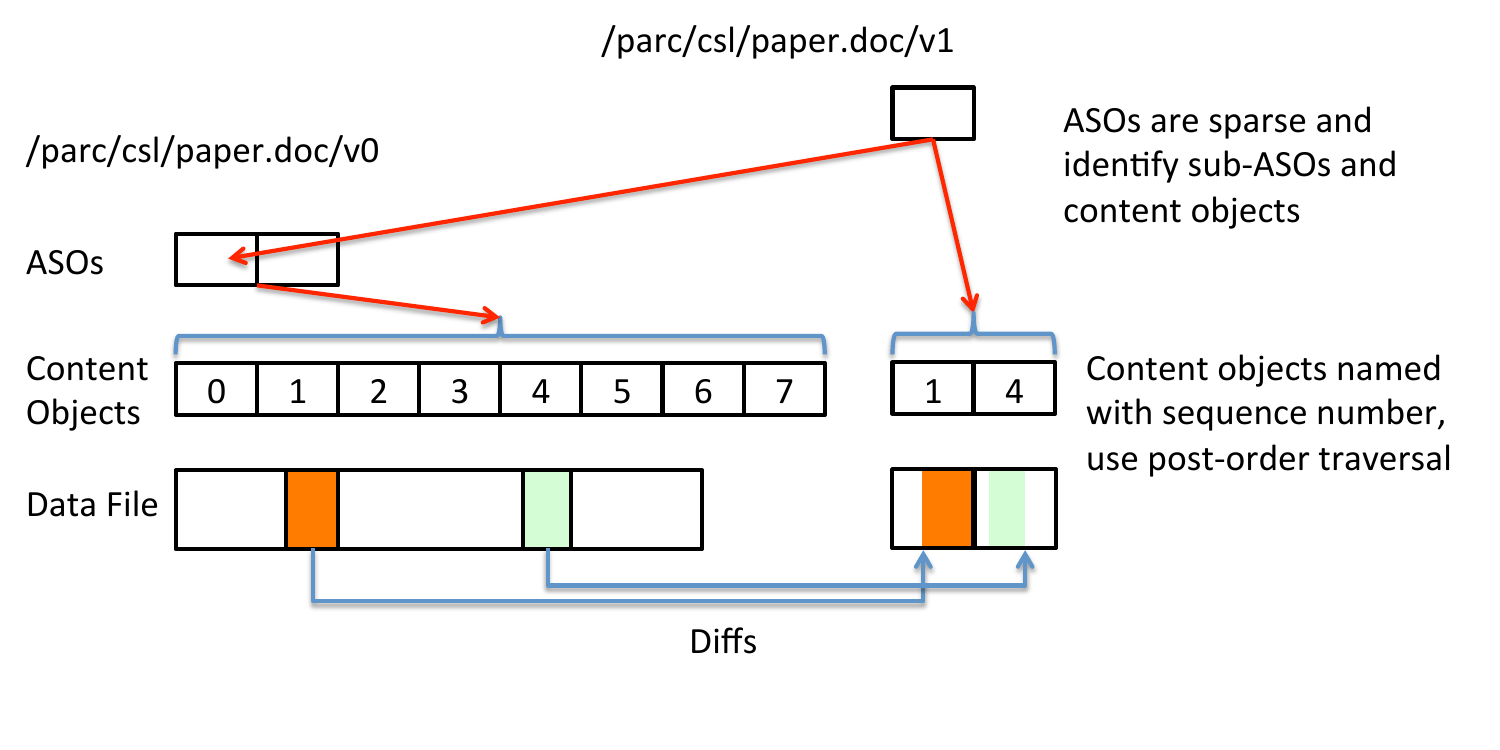}
\caption{Differential data using content object diffs}
\label{fig:objectdiffs}
\end{figure}

In this variation, instead of doing binary diffs on the data file and encoding that diff,
this variation uses diffs on the content object sequence numbers.  Fig.~\ref{fig:objectdiffs}
shows an example where the original data files encodes to 8 content objects and
the diff to version 1 replaces two of those content objects with new objects.

To reconstruct a version, one does a post-order traversal of the secure catalog and
only uses the right-most occurrence of a content object sequence number.

As with the previous variation, this method may incur high overhead of
bytes are inserted, causing a right-shift of content objects.  If bytes are
removed, one could easily elide those bytes using a single content object,
which may be empty if all bytes in the previous version of that object are replaced.

In this variation, because the diff is done at the content object level, each
content object must contain all the bytes being replaced in the previous
content objects.  As such, if a small number of bytes are changed, for example 128 bytes
in an 8KB content object, all 8KB of the new version must ben encoded
in the replacement content object.

\section{Byte-offset Content Object diffs}
\begin{figure}
\centering
\includegraphics[width=\linewidth, clip, page=3, trim=0in 0in 0in 2.5in]{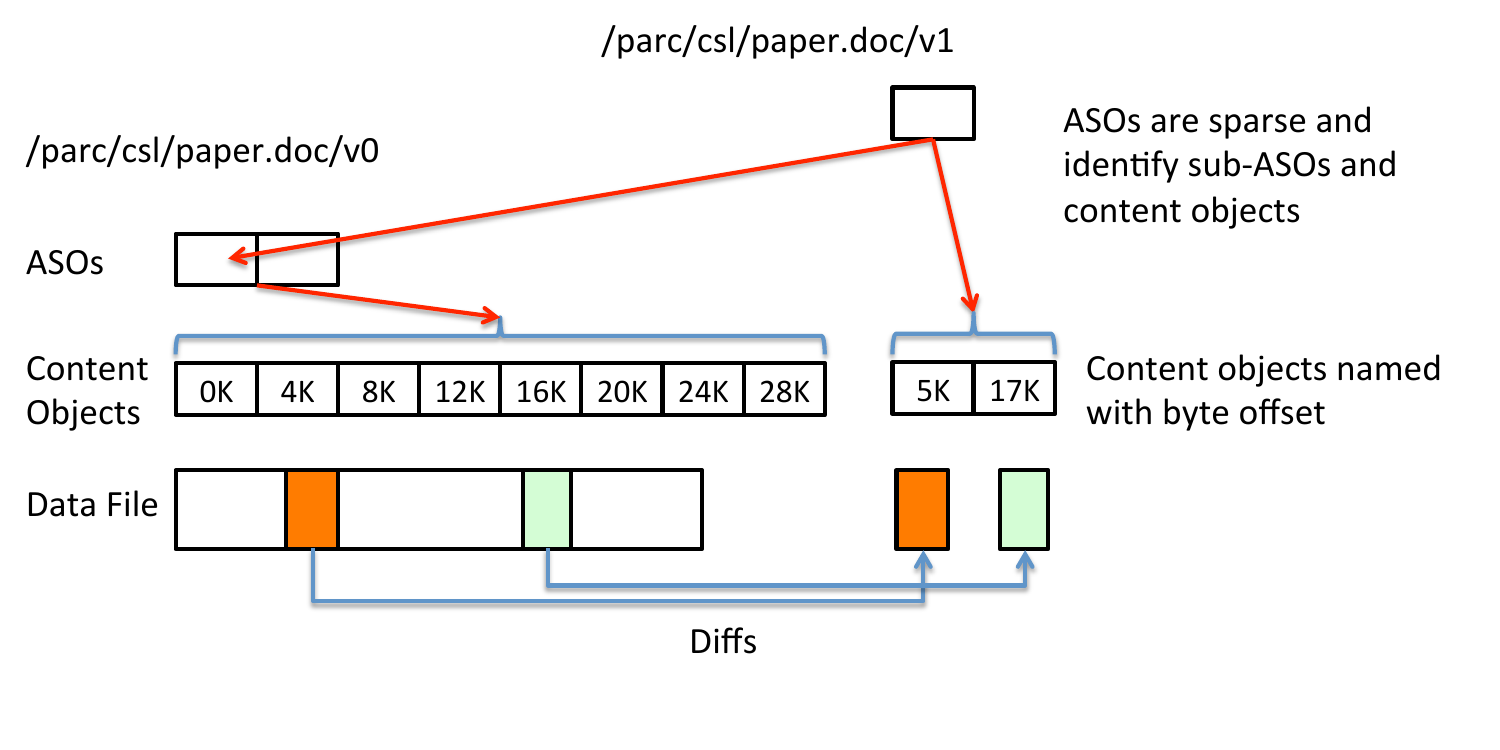}
\caption{Differential data with byte offsets}
\label{fig:bytediffs}
\end{figure}

In this variation, shown in Fig.~\ref{fig:bytediffs}, uses byte offsets to label where
the bytes inside a content object should be placed in the previous version.
The payload of the content object indicates if it represents an ``insert'' or
''replace'' or ''deletion'' operation.

To reconstruct a version, one does a post-order traversal of the secure
catalog tree and maintains an interval graph of the data file.  

\section{Secure Catalog for Chunk Enumeration}
\begin{figure}
\centering
\includegraphics[width=\linewidth, clip, page=4, trim=0in 0in 0in 2.5in]{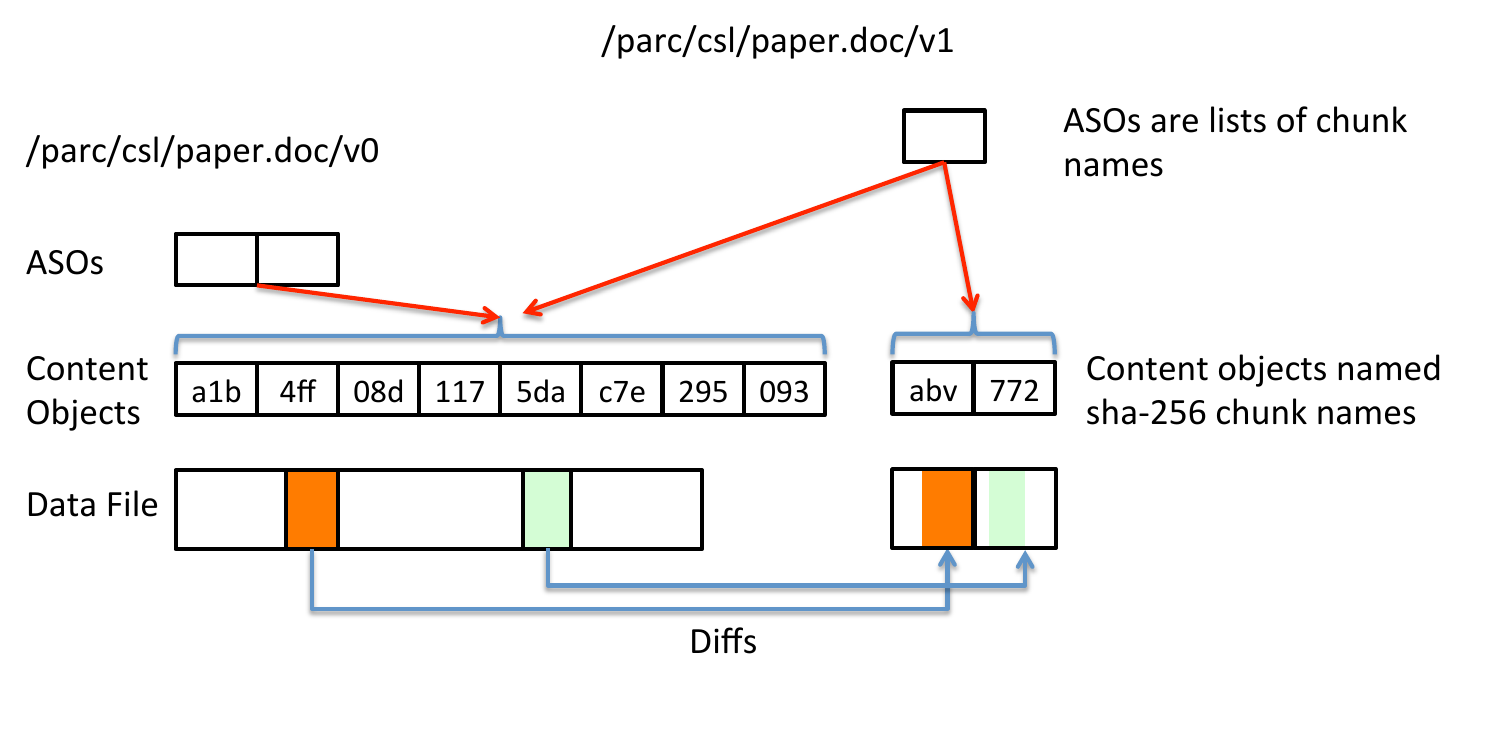}
\caption{Chunk de-duplication encoding}
\label{fig:chunks}
\end{figure}

In this variation, depicted in Fig.~\ref{fig:chunks}, a data file is broken up in to chunks, 
using data de-duplication technology.
Chunks usually vary from 4KB to 16KB, depending on the data and the technology used.

Each chunk is named by its cryptographic hash, such as its SHA-256 name.  In one variation,
the names appear similar to \url{/parc/csl/paper.doc/<chunk_hash>}, where \url{<chunk_hash>}
is the 32-byte hash.  In another variation, chunks may be kept under a higher-level chunk
repository, such as \url{/parc/<chunk_hash>}.

The secure catalog has a versioned name, such as \url{/parc/csl/paper.doc/v0}, where \url{v0} 
indicates it is the first version.  The secure catalog then enumerates all chunk hashes in 
order.  The secure catalog only needs to name the hashes to the extent needed to find them.  
For example, if the system stores the chunks under \url{/parc/csl/papers.doc}, the secure catalog
only needs to state this in one place, then the remainder of the entries are only the 32-byte chunk
names.

For the next version, such as \url{/parc/csl/papers.doc/v1}, the new secure catalog enumerates
the new set of chunks.

\section{Diff-based Secure Catalogs for Chunk Enumeration}
\label{sec:diffchunks}

\begin{figure}
\centering
\includegraphics[width=\linewidth, clip, page=5, trim=0in 0in 0in 2.5in]{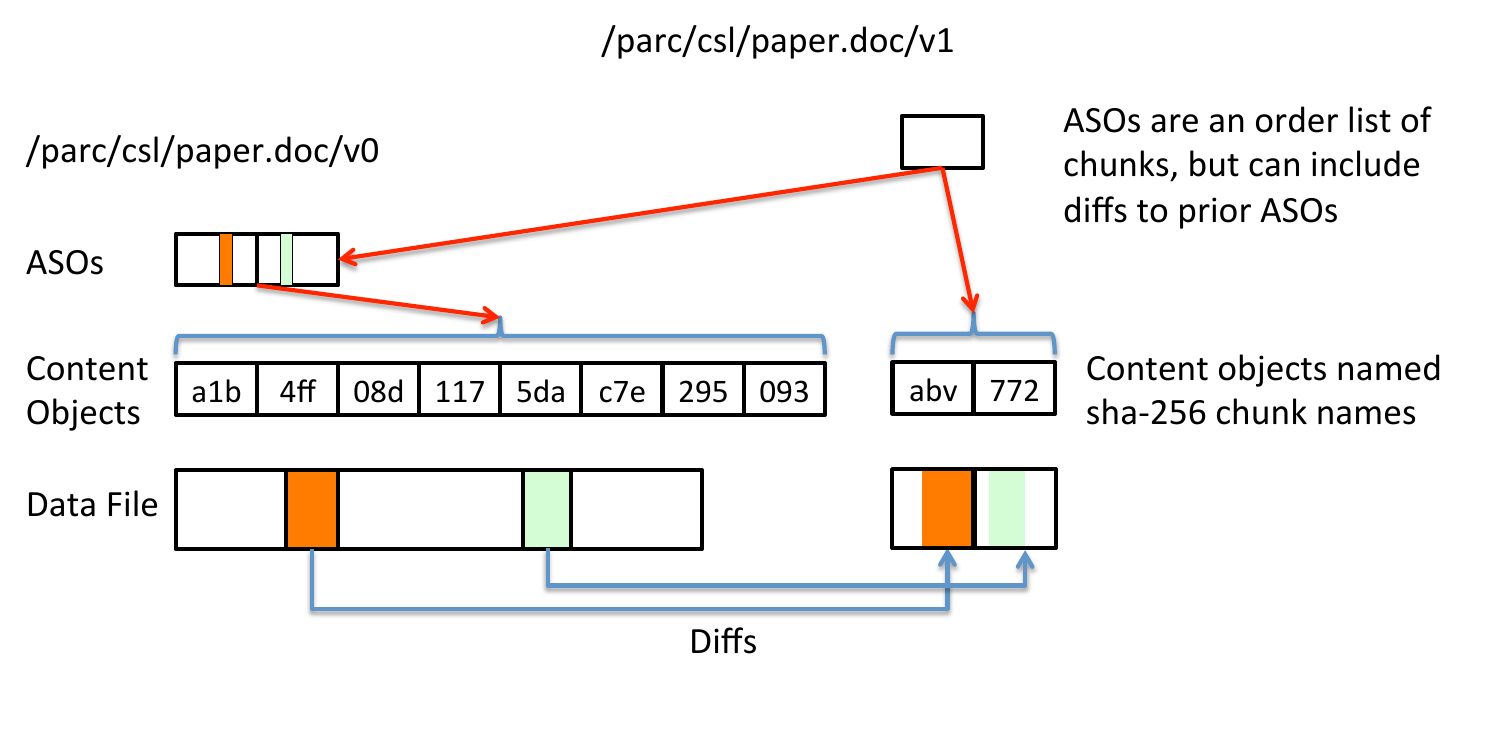}
\caption{Chunk de-duplication encoding with ASO diffs}
\label{fig:diffchunks}
\end{figure}

This variation is also based on chunks, as shown in Fig.~\ref{fig:diffchunks}.  
In this case, the secure catalog is a diff of previous secure catalogs.  This allows
easy insertion or removal of chunks and because of the differential encoding, the
secure catalogs for subsequent versions may be small.  

Diffs do not need to be solely to the previous version's catalog.  The secure
catalog diff may point to any number of earlier versions indicating the diffs
to those versions.  The final secure catalog is constructed via a post-order
traversal of the secure catalog tree.

\section{Conclusion}
Diff Based Content Networking offers several efficient ways to encode data
over versions that do not require transmitting all the bytes of an object with
each version.  Furthermore, it leverages perviously-signed secure catalogs
to reduce the signing overhead of publishing a new version.



\bibliographystyle{unsrt}
\bibliography{dcn}


\end{document}